\documentclass[review]{elsarticle}

\usepackage{hyperref}
\usepackage{xcolor}

\journal{Icarus}

\biboptions{authoryear}

\begin{document}

\begin{frontmatter}

\title{Seismic response of the Mars Curiosity Rover:  Implications for future planetary seismology}

\author[jpl]{Mark P. Panning\corref{corauthor}}
\cortext[corauthor]{Corresponding author}
\ead{Mark.P.Panning@jpl.nasa.gov}

\author[jpl]{Sharon Kedar}
\ead{Sharon.Kedar@jpl.nasa.gov}

\address[jpl]{Jet Propulsion Laboratory, California Institute of Technology, 4800 Oak Grove Drive, Pasadena, CA 91109, USA}

\begin{abstract}
Seismic measurements are an important tool for exploration of planetary interiors, but may not be included in missions due to perceived complexity in placement of sensitive instruments on the surface.  To help address this concern, we assess the fidelity of recordings of ground motion by an instrument placed on the deck of the engineering model of the Mars Science Laboratory compared with an identical instrument placed on the ground directly beneath.  Comparison of the recordings reveals clear recordings of teleseismic earthquakes on both instruments.  The transfer function between the instruments demonstrates the deck instrument is affected by resonance frequencies of the lander, and does not faithfully record ground motion at these frequencies or higher.  In addition, additional decoherence is observed near 1 Hz during periods of strong airflow due to air conditioning cycling.  However, excellent coherence and a transfer function near 1 can be observed in the important seismic band between 2 and 30 seconds at all times and extending up to the lander resonances during the night time when air conditioning was not running.  This suggests a deck-mounted seismic instrument may be able to provide valuable science return without requiring additional deployment complexity.
\end{abstract}


\end{frontmatter}


\section{Introduction}
In the coming decades NASA is expected to launch multiple missions to explore the Ocean Worlds of Jupiter and Saturn.  In addition, landed mission concepts to study rocky bodies (the Moon, Mars, Venus and asteroids) are routinely developed and proposed.  Seismic instruments will likely be a common instrument to many such landed mission concepts, such as the proposed Europa Lander and Dragonfly missions \citep[e.g.][]{Hand+2017,Turtle+2018}, as they are the most efficient and proven tool to explore a planetary body's interior.  Yet, since 1976 (Viking 2) no seismometer has been included in any of NASA's lander or rover missions, and the last NASA seismometer to operate outside Earth was turned off more than forty years ago (Apollo Passive Seismic Experiment, September 1977).  The InSight mission, slated for launch in 2018, will be the first since Viking to use seismometers to learn about the interior of Mars.

The most commonly cited reason for this deficit in this key geophysical data set is the complexities involved in launching, landing, and emplacing highly-sensitive and delicate seismometers on extra-terrestrial bodies.  Seismic networks requiring multiple landers are deemed even more costly, impractical or risky (perhaps with the exception of a Lunar Geophysical Network, as prioritized in the planetary science decadal survey \citep{Decadal2011}).  More specifically, it is a common perception that direct coupling to the planetary surface is a must.  This requirement drives complex robotic (InSight) or human (Apollo) emplacement of the seismometers on the planetary surface, and results in increased mission complexity, risk, and cost.

\begin{figure}
\includegraphics[width=13pc]{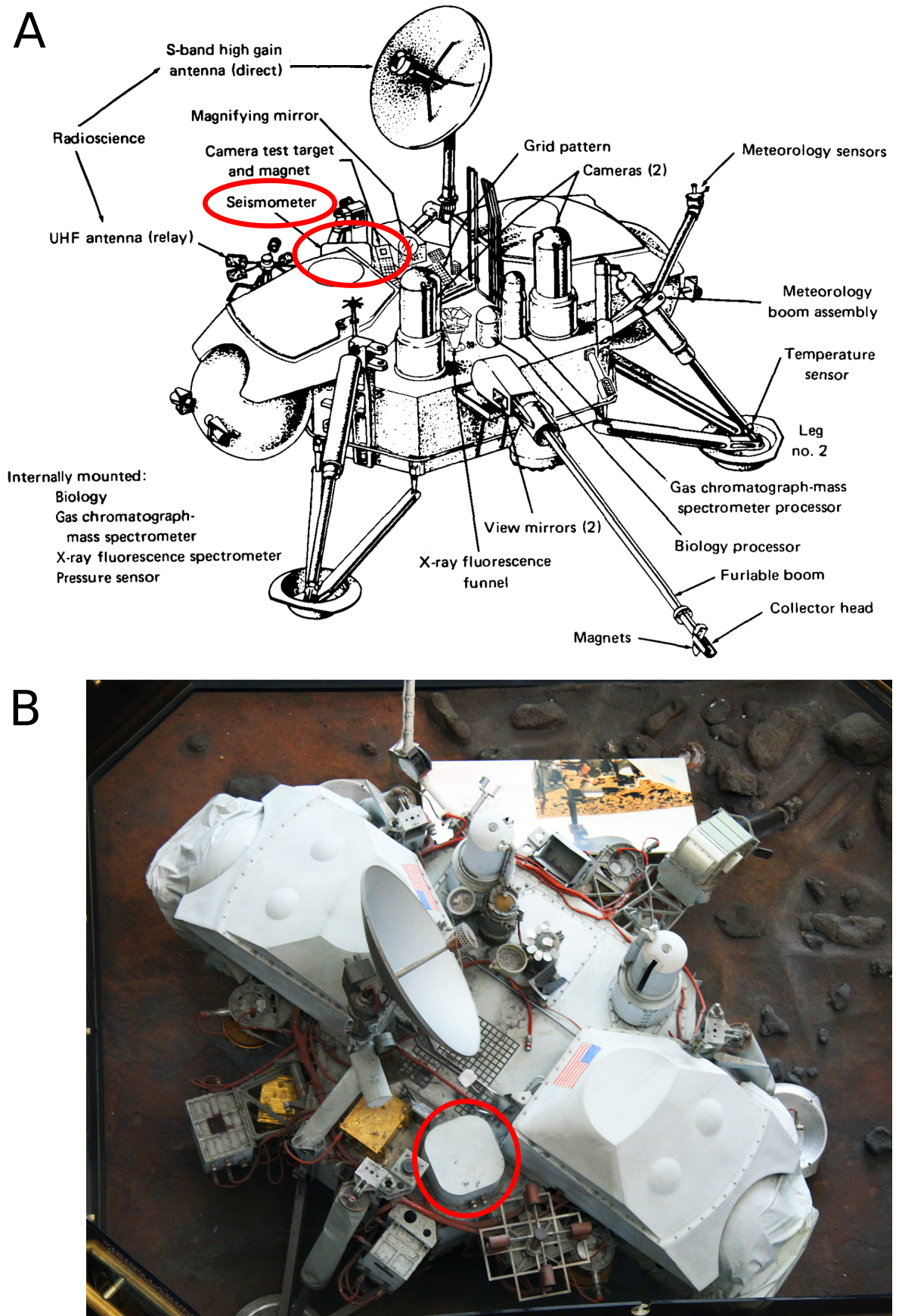}
\caption{(A) Landed science configuration of the Viking landers \citep{NASA1976} with seismometer deck location highlighted in red.  (B) Overhead photo (taken by Tim Evanson) of Viking lander backup at the Smithsonian Air and Space Museum in Washington, D.C., with seismometer circled in red.}
\label{vikingfig}
\end{figure}

There is no dispute that careful coupling to the ground is highly desirable and yields superior performance in terms of sensitivity and environmental noise, especially in atmosphere bearing planets.  The Viking seismometer, which remained exposed to the Martian wind (Figure~\ref{vikingfig}), is habitually mentioned as an example in which lack of coupling to the Martian surface resulted in a severely increased noise environment and consequently a reduced sensitivity that may have contributed to the absence of Mars quakes from the Viking seismic record \citep{Anderson+1977}.  There were other important factors, however, that contributed to the lack of clear detection of seismic events.  The instruments launched on the Viking landers were short period accelerometer instruments with sensitivity strongly peaked for frequencies near 3 Hz.  Even near the peak sensitivity at 3 Hz, the instrument sensitivity was worse than the widely used (at the time) World-Wide Standard Seismograph Network (WWSSN) short-period instrument by a factor of $\sim$2, and by roughly an order of magnitude at 1 Hz and below, while the Apollo instruments were even more sensitive than that \citep{Anderson+1977}.  Finally, due to data constraints, most of the data returned from the instrument were sent in a compressed `event mode' rather than the raw recorded data, which consisted of envelope amplitude recorded at $\sim$1 Hz as well as the number of positive-going zero crossings \citep{Lorenz+2017b}.  Such a format precludes most tools for the processing of modern digital seismograms to separate signal from noise.  Yet, in the forty years since Viking, during which seismic sensing and digital recording have remarkably improved, no attempt was made to quantitatively or experimentally explore whether modern seismometers operating on a lander deck can, despite lander response to ground motion and wind, produce quality seismic data of sufficient quality to study the interiors of previously unexplored planetary bodies.

To fully answer this question, one would need to address the specific science requirements of the mission in question.  In addition to the obvious varying operational requirements (gravity, temperature, radiation environment, etc), the dramatic differences between planets or moons also imply a wide range of seismic activity \citep[e.g.][]{,Golombek+1992,Panning+2018}, which invariably dictates different seismic noise floor requirements.  In turn, the seismic noise environment is comprised of a multitude of contributors that in addition to planet's expected tectonic and non-tectonic seismicity is strongly affected by the spacecraft that carries the seismometer to the surface \citep{Murdoch+2017b,Murdoch+2017a}.  We can consider the total noise budget to be a combination of the instrument noise, related to the details of the instrument design, and environmental and installation noise caused by both spacecraft activities and the imperfect coupling with ground motion caused by any structural elements between the ground and the seismometer.  Therefore, a key question faced by future landed mission is whether or not the seismic response of the spacecraft enhances undesired noise contributors to the point that it renders a seismometer, as sensitive as it may be, incapable of delivering meaningful geophysical science, unless it is firmly coupled to the ground and is protected from the elements.  In this paper we take the first step towards answering this question, by isolating the seismic response of the Mars Science Laboratory ``Curiosity'' rover \citep{Grotzinger+2012}.  

\section{Experimental setup}
The Mars Science Laboratory (MSL) has been in operation on the surface of Mars since its landing on August 5, 2012.  As part of the routine operation of the MSL rover on the Martian surface, an identical replica (known as an Engineering Model) is operated at the Jet Propulsion Laboratory (JPL) ``Mars Yard''.  The replica rover is used to test and exercise maneuvers on Earth before commands are beamed to the rover on Mars.  The MSL rover is the biggest, most complex robotic spacecraft to ever land on an extraterrestrial body, and so provides a ``worst case'' scenario for a lander-induced seismic noise, compared to a small lander or a penetrometer whose natural resonant frequencies are likely to be well outside the seismic frequency band of interest.

\begin{figure}
\includegraphics[width=15pc]{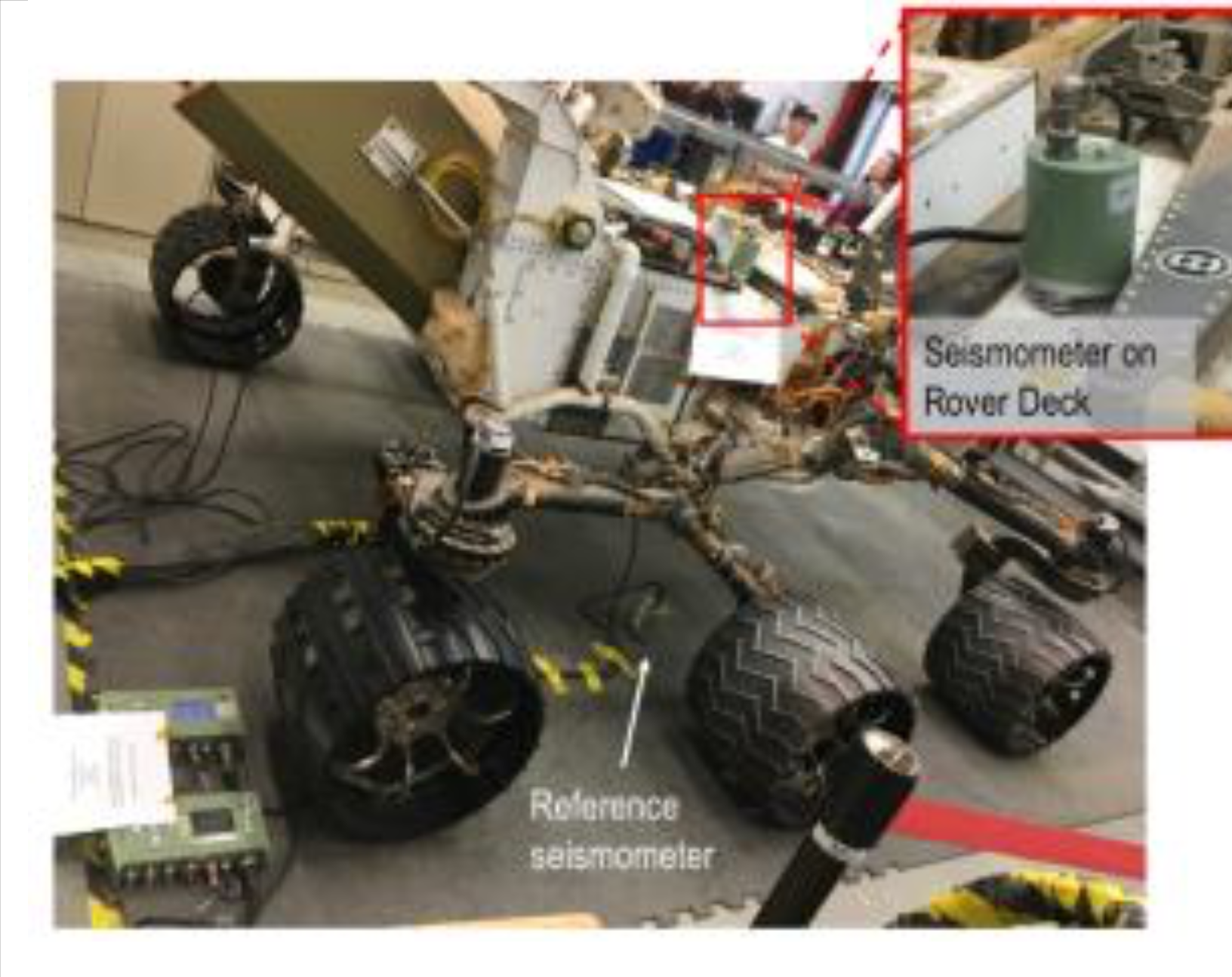}
\caption{Photo of testbed MSL ``Curiosity'' rover in Mars Yard hangar with reference and deck seismometer locations highlighted.}
\label{roverfig}
\end{figure}

In order to isolate the MSL rover seismic response, we used two Trillium Compact seismometers.  One was placed on the replica MSL rover deck and one was placed directly underneath it for reference (Figure~\ref{roverfig}).  The study was conducted inside the Mars Yard hanger, which isolated the rover and seismometers from external wind and diurnal temperature variations.  It was, however, exposed to mechanical and cultural noise induced by air conditioning, people and vehicles.  To minimize the latter we carried out the measurement over a weekend, when cultural noise is reduced.  By minimizing other sources of noise, we attempt to isolate how the recovered seismic signal is affected simply due to placement on the deck of a landed science mission rather than being more directly coupled to the ground.

\section{Data analysis}
\begin{figure}
\includegraphics[width=15pc]{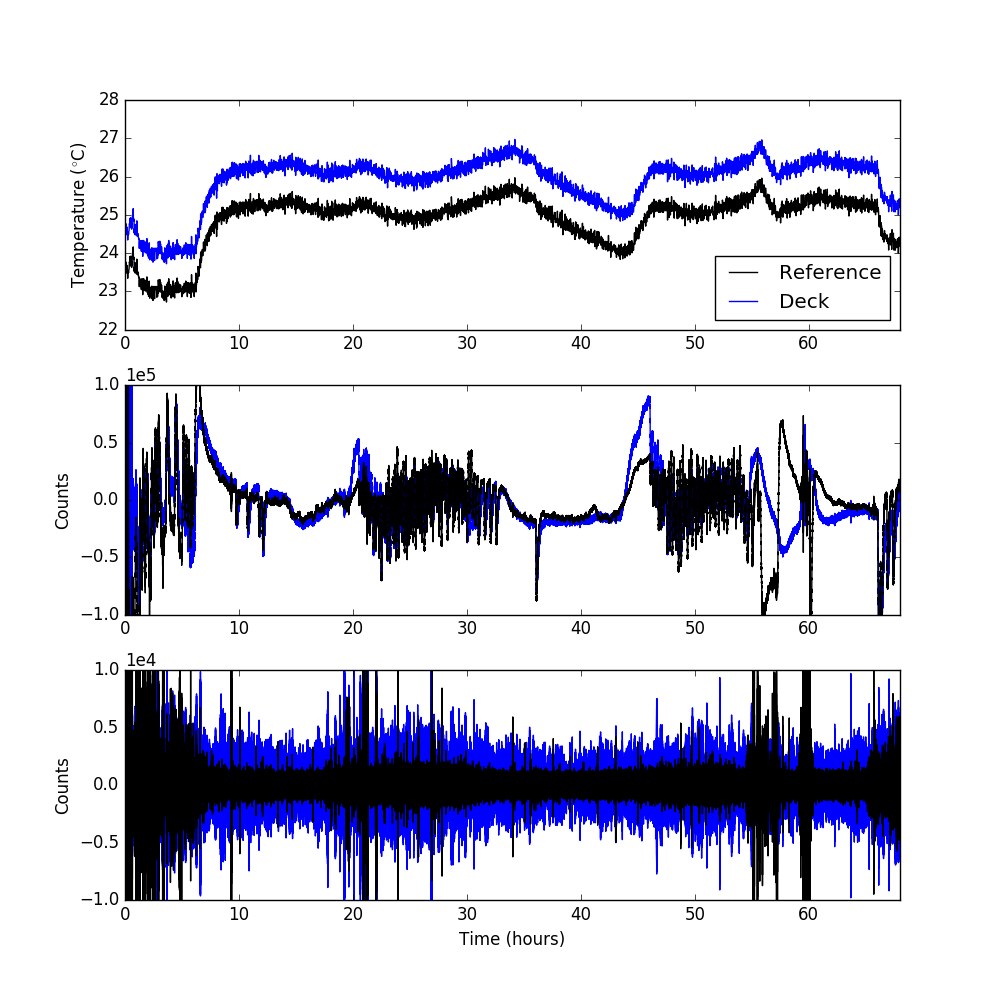}
\caption{Temperature records (top) for the entire experiment duration for the reference (black) and deck (blue) seismometers with the deck seismometer temperature offset up by 1$^{\circ}$C for clarity.  Unfiltered seismic records (middle) and seismic records highpass filtered above 1 Hz (bottom) are also shown.}
\label{tempfig}
\end{figure}

\begin{figure}
\includegraphics[width=20pc]{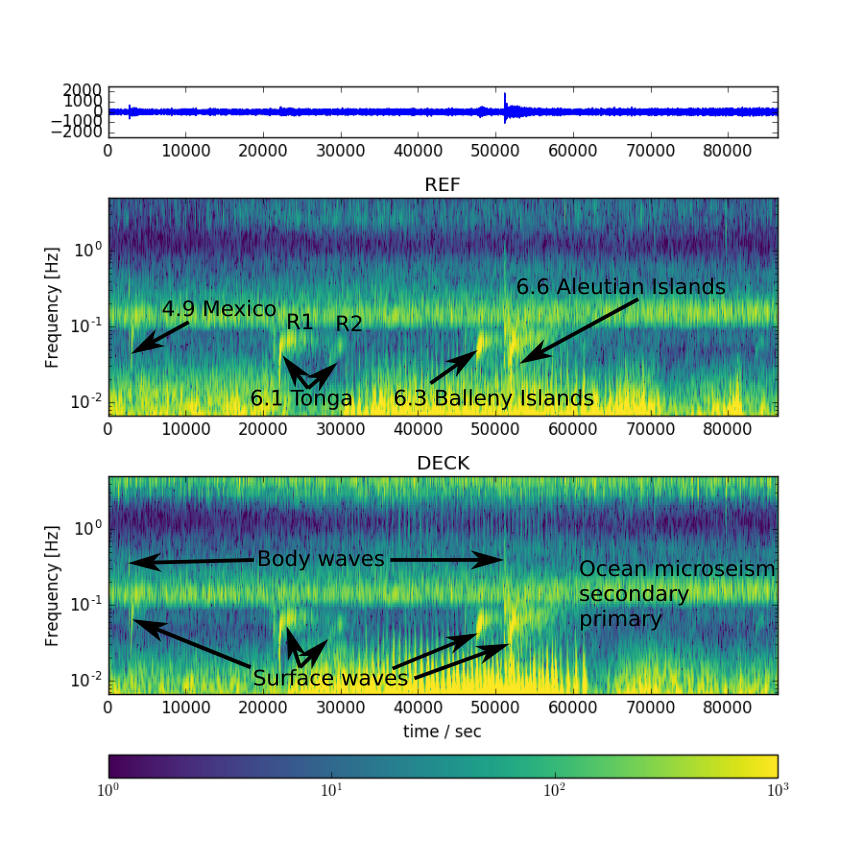}
\caption{Spectrograms for the vertical components of the reference (REF, middle) and deck seismometer (DECK, bottom), with the deck seismogram filtered between 0.05 and 2 Hz shown at the top in blue.  Color scale is defined by amplitude of the continuous wavelet transform \citep{Kristekova+2009} as a function of time and frequency.  Records are for a 24 hour period beginning November 8, 2017 at 08:30:00 UTC.}
\label{specfig}
\end{figure}

\begin{figure}
\includegraphics[width=15pc]{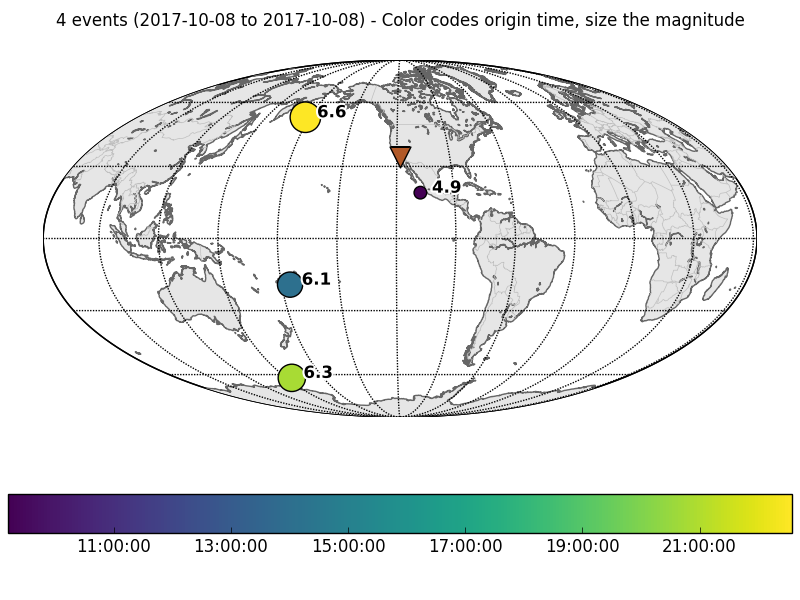}
\caption{Events identified in spectrograms in figure~\ref{specfig} are plotted as circles with size scaled to magnitude and color scaled to origin time according to color bar at bottom.  All events greater than magnitude 6 globally in the NEIC catalog during this time are plotted, as well as all events greater than magnitude 4.5 within 30$^{\circ}$ of the recording location (shown as inverted triangle).}
\label{eventsfig}
\end{figure}

The comparison of the raw records of the seismogram on the MSL rover deck and the one on the ground below (Figure~\ref{tempfig}) reveals a record dominated by very long period signals which often show up on both records, but sometimes deviate significantly. There is also a clear diurnal pattern related to cultural effects with higher noise with different frequency characteristics during the day.  When the long period signals are filtered out, the records become dominated by high frequency oscillations that do not closely correlate between the sensors, with the sensor on the deck generally seeing higher amplitude signals.  However, a closer look at spectrograms of the data (Figure~\ref{specfig} for a select window, and for all data in supplementary material) reveals that more useful and coherent data can be seen across a broad frequency band between these extremes.  Spectrograms are calculated using a continuous wavelet transform \citep{Kristekova+2009} as implemented in the python seismology package obspy \citep{Krischer+2015}.  Both records show clear, consistent recordings of the secondary microseismic band, with a less prominent primary microseism between 0.05 and 0.2 Hz (periods from 5 to 20 seconds) \citep[e.g.][]{Longuet-Higgins1950}, indicating the instrumental and installation noise levels of both instruments are below the ambient ground motion.  Figure~\ref{specfig} also shows signals of teleseismic waves from 3 large global earthquakes greater than magnitude 6, and one smaller event offshore from Mexico (locations shown in Figure~\ref{eventsfig}).   These events can be seen most prominently at periods between 10 and 30 seconds where high amplitude Rayleigh wave energy arrives with a velocity dispersion visible as repeated curved features in the spectrograms, but there is also a strong broadband signal to higher frequencies up to 1 Hz associated with the body wave arrivals from the largest event in the time window, as well as the smaller, closer event.  For the event in Tonga, the signal from the Rayleigh wave that traveled the longer way around the planet (R2) can even be identified as it arrives during a quiet period during the nighttime.  The consistency of observations of ground motion linked to planetary activity between the deck and ground-placed instruments across the important seismic frequency band between 2 and 30 seconds indicates that deck deployment of sufficiently sensitive seismometers may be a viable option that should be further explored for future landed missions.

\subsection{MSL Lander Response}
All seismic instruments have a sensitivity to ground motion that is frequency dependent, and within limits defined by the instrument design can be considered as a linear function described by phase and amplitude shifts.  This conversion between ground motion and seismometer output is often described by a linear multiplication in the frequency domain by a transfer function \citep[e.g.][]{Aki+2002}.  Many techniques are used to characterize the transfer function of a particular instrument, but one common approach is to simply install the instrument in the same fashion as very well-calibrated instrument and compare the output data of the two instruments in the frequency domain. Our experimental setup with identical instruments installed on the lander deck and on the ground below lets us isolate the contribution to the transfer function due to installation on the lander deck rather than directly coupled with the ground. 

\begin{figure}
\includegraphics[width=15pc]{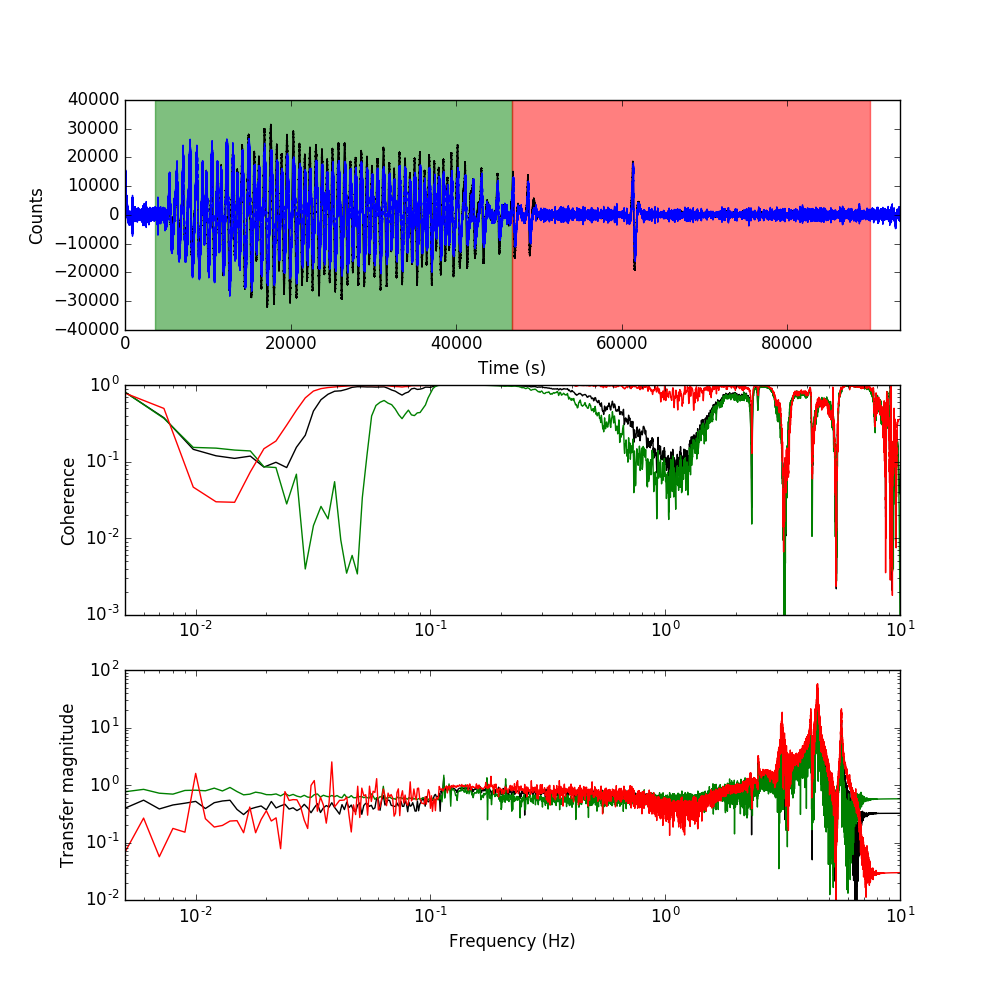}
\caption{Example vertical component seismograms (top) filtered between 0.001 and 5 Hz with reference seismometer plotted in black and deck in blue.  The green window highlights a typical daytime period dominated by relatively long-period ($\sim$5 minutes) signals due to air-conditioning cycles, while the red window is during the nighttime.  Coherence between the deck and reference seismograms (middle) and transfer function from reference to deck (bottom) are also plotted.  In both coherence and transfer function, the green line corresponds to the calculation over the daytime window, while the red one is over the nighttime window, while the black one is over both time windows.}
\label{compfig}
\end{figure}

There are 2 related frequency-dependent quantities to consider when trying to determine the true transfer function from an experimental setup like this, the coherency and estimated transfer function (Figure~\ref{compfig}).  First, we need to determine whether the two signals (in this case the reference seismometer and the deck seismometer, which we can call $x(t)$ and $y(t)$) can be linearly related.  This can be assessed with the magnitude squared coherence $C_{xy}$ \citep[e.g.][]{Stoica+2005},
\begin{equation}
C_{xy}(f) = \frac{\left|P_{xy}(f)\right|^2}{P_{xx}(f)P_{yy}(f)},
\end{equation}
where $P_{xy}$ is the cross-spectral density and $P_{xx}$ and $P_{yy}$ are the power spectral densities of the two signals as a function of frequency $f$.  This value ranges between 0 and 1, and is a measure of the fractional power of the output signal $y$ that can be produced by the input signal $x$.  For this calculation, power spectral densities are estimated with the Welch method \citep{Welch1967} as implemented in the python package scipy (\url{http://www.scipy.org/}).  In frequency bands where coherence is high, it is possible to define a transfer function to recover the reference seismogram (and therefore the true ground motion under the assumption that the reference seismometer is accurately recording ground motion) from the deck seismogram, while frequency bands with low coherence can not be used to reconstruct actual ground motion.  If the coherence is close to 1, the transfer function in theory could be determined directly by spectral division, $T(f) = Y(f)/X(f)$, where $X(f)$ and $Y(f)$ are the Fourier transforms of  $x(t)$ and $y(t)$.  In practice, however, such a calculation is numerically unstable in the presence of portions of $X(f)$ near zero.  We choose to calculate transfer functions using Tikhonov regularization, 
\begin{equation}
T(f) = \frac{X^*(f)Y(f)}{(X^*(f)X(f)+\lambda)},
\end{equation}
where $X^*(f)$ is the complex conjugate of $X(f)$ and $\lambda$ is a regularization coefficient, and we make our final transfer function estimate by averaging over multiple windows of length $\tau$.  In general, increasing $\lambda$ or increasing the number of windows used to calculate the average transfer function will lead to a smoother estimate, but will lead to a less accurate reconstruction of $Y(f)$ by the product $X(f)T(f)$.  For the transfer functions shown in figure~\ref{compfig}, we used $\lambda=10^6$ and divided the record into 1000 second windows for the averaging.

At frequencies above 2 Hz, coherence is very low between the instruments and the transfer function consists of several strong magnification peaks.  These are associated with resonant frequencies of the rover structure, and oscillations of the rover dominate the signal at these frequencies.  A deck-installed seismometer will not be able to observe true ground motion in these frequency bands, but the precise resonant frequencies depend on the size and construction details of the particular landed spacecraft.  Given that the MSL rover is the largest planetary landed science platform to date, though, most other landers and rovers will have higher resonant frequencies, increasing the useful bandwidth for seismic recording.  

Another pronounced dip in coherence is observed near 1 Hz, although this is not matched by a corresponding magnification peak in the transfer function.  This loss of coherence is dominantly caused by episodic high noise periods in the data that appear to be time periods during the day when the air conditioning in the hangar was active.  This is clear as the coherence during the night does not show this dip near 1 Hz.  Airflow across the seismometer is an environmental noise source that is frequently a concern for surface installations, which is one reason why the InSight seismic instrument includes a wind and thermal shield \citep[e.g.][]{Murdoch+2017a}, as wind was the dominant noise source for the Viking instrument \citep{Anderson+1977,Lorenz+2017b}.  In this particular experiment, the wind noise during active air conditioning was incoherent between the reference and deck-installed seismometers, but only affected the data near 1 Hz.  From frequencies between 0.04 and 0.5 Hz (periods from 2 to 25 seconds), however, the instruments have high coherence and a transfer function near 1.  This indicates that the deck-installed seismometer recorded an essentially identical signature to the ground-installed reference seismometer across this critical seismic frequency band over the duration of this experiment.

\begin{figure}
\includegraphics[width=14pc]{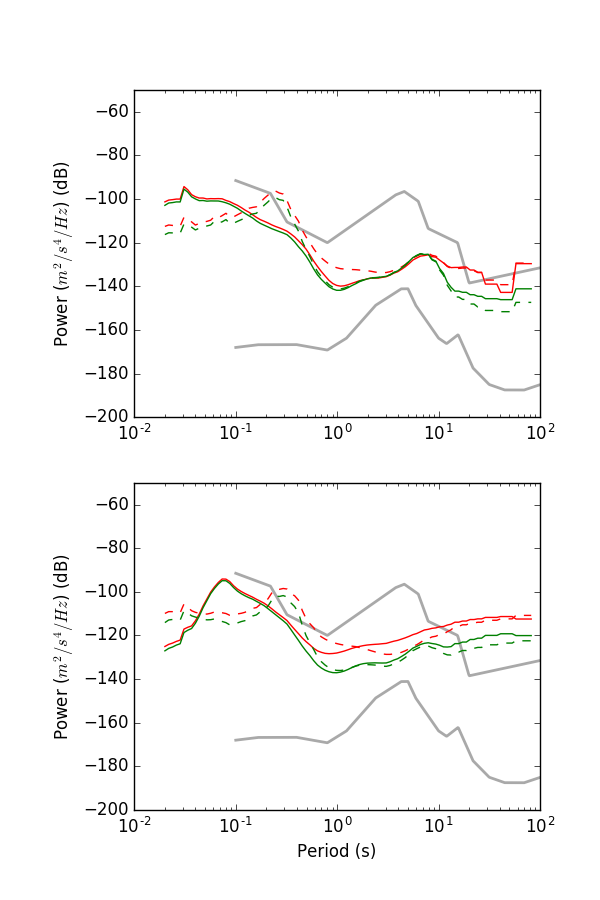}
\caption{Mean power spectral density estimations for the vertical component (top) and averaged horizontal components (bottom).  Daytime power is shown by red lines, while nighttime is shown with green lines.  In both panels, the signal power for the reference seismometer is shown with a solid line, while the deck seismometer is shown with dashed lines.  For reference the high and low noise Earth noise models \citep{Peterson1993} are shown with gray lines.}
\label{psd_summary_fig}
\end{figure}

Another way to visualize the difference between the 2 seismometers and the difference between the day and night records is to examine the power spectral density (PSD) of the signals over the duration of the experiment (figure~\ref{psd_summary_fig}).  To calculate this, we use a probabilistic approach to determine the PSD that is commonly used when assessing noise levels recorded by seismic stations \citep{McNamara+2004b}, as implemented in probabilisitic power spectral density (PPSD) tool in the signal processing toolkit of ObsPy \citep{Krischer+2015}.  For this application, we divided the seismic records into overlapping 10 minutes segments, and estimate the power spectral density for each segment.  The resulting estimates are stacked into histograms to estimate the probability density function of the PSD.  This is done to avoid high amplitude portions of the data (such as those caused by discrete seismic events) dominating the PSD estimate.  In figure~\ref{psd_summary_fig}, we plot the mean values of these probability density functions.

The PSD estimates clearly demonstrate similarities and differences between day and night and between the reference and deck seismometers.  At frequencies above 1 Hz, we can clearly see the signal magnification of $\sim$20 dB at a period of 0.3-0.5 seconds (2-3 Hz) in the deck seismometer (dashed lines in figue~\ref{psd_summary_fig}) due to the resonance of the rover in both the vertical (top) and averaged horizontal components (bottom).  At periods less than 0.1 second (greater than 10 Hz), on the other hand, the signal shows greater power in the reference seismometer, indicating that the rover body acts as a shock absorber above its resonance frequencies.  Near 1 Hz, the nighttime PSD estimates (green lines) are similar for both seismometers, consistent with the high coherence and transfer function near 1 shown in figure~\ref{compfig}, but in the daytime estimates (red), the power on the deck is higher by $\sim$10 dB, which demonstrates the greater noise on the deck due to air conditioning airflow.  At long periods greater than 10 s, there is little difference between the deck and reference seismometers, but there is a clear day/night separation, with greater noise during the day.  The horizontal components are consistently noisier than the vertical component for periods longer than 1 s (frequency less than 1 Hz), and show a $\sim$10 dB offset between day and night records.  The prominent microseismic peak near 10 s which can be seen in the vertical component PSD is only barely visible in the horizontal component PSD during the night time, and lost in other sources of noise during the day.  This increased horizontal noise, however is present in both the deck and reference seismometers suggesting the greater horizontal noise here cannot be attributed to deck deployment, but is instead reflective of other noise sources.

\section{Discussion and Conclusions}
The equivalence of performance between the deck and reference seismometers across an important seismic frequency band suggests that it may be possible to obtain valuable seismic data in future landed science missions on planetary bodies without necessarily requiring significant and complicated spacecraft accommodation for surface installation. Obviously, careful modeling of the amplitudes of actual expected signatures in the frequency bands not affected by lander resonances on target planetary bodies would be required to maximize expected science returns for any particular mission, but the challenge in instrument design is effectively the same in this frequency range whether the instrument is mounted on the deck or on the surface of the planet.  

This study was also performed during a time when the MSL rover was not in use, and therefore specifically does not address noise generated by spacecraft operations.  Such noise will undoubtedly be more pronounced for an instrument mounted on a deck than on the ground near a lander or a rover.  However, most missions have significant quiet time periods where no active operations are ongoing, and a seismometer could passively record data during these time periods.

\section*{Acknowledgments}
\begin{sloppypar}
This work was supported by strategic research and technology funds from the Jet Propulsion Laboratory, California Institute of Technology, under a contract with the National Aeronautics and Space Administration.  Raw data is available at \url{http://www.github.com/mpanning/MSL-experiment}.  Trillium Compact instruments were provided by Robert Clayton.  Image of Viking lander backup at Smithsonian Air and Space Museum taken by Tim Evanson was used under a creative commons license (\url{https://creativecommons.org/licenses/by-sa/2.0/}) and was modified to highlight seismometer location.  \copyright~2018 California Institute of Technology.  Government sponsorship acknowledged.
\end{sloppypar}

\bibliography{biblio}

\end{document}